\documentclass[journal=jacsat,manuscript=article]{achemso}

\usepackage{chemformula} 
\usepackage[T1]{fontenc} 
\usepackage{textcomp}
\usepackage{graphicx}
\usepackage{caption}    
\usepackage[english]{babel}
\usepackage[utf8]{inputenc} 
\usepackage{siunitx}
\usepackage{comment}
\usepackage{ulem}
\usepackage{amsmath,amssymb}

 \author{Marina Cagnon Trouche}
 \affiliation{%
 Laboratoire de Physique de l’Ecole normale supérieure, ENS, Université PSL, CNRS, Sorbonne Université, Université Paris Cité, F-75005 Paris, France\\
}%
\altaffiliation{These authors contributed equally to this work.}

\author{Ernest Ruby}
\affiliation{%
 Université Paris-Saclay, ENS Paris-Saclay, Centrale Supélec, CNRS, LuMIn (UMR 9024), Gif-sur-Yvette, France\\
}%
\altaffiliation{These authors contributed equally to this work.}

\author{Margaux Cartier}
\affiliation{%
 Laboratoire de Physique de l’Ecole normale supérieure, ENS, Université PSL, CNRS, Sorbonne Université, Université Paris Cité, F-75005 Paris, France\\
}%

\author{Christophe Voisin}%
\affiliation{%
Laboratoire de Physique de l’Ecole normale supérieure, ENS, Université PSL, CNRS, Sorbonne Université, Université Paris Cité, F-75005 Paris, France\\
}%

\author{Maxime Vallet}%
\affiliation{%
Université Paris-Saclay, CentraleSupélec, CNRS, Laboratoire SPMS (UMR8580), 91190 Gif-sur-Yvette, France\\
}%
\alsoaffiliation{Université Paris-Saclay, ENS Paris-Saclay, CentraleSupélec, CNRS, LMPS – Laboratoire de Mécanique Paris-Saclay (UMR 9026), 91190 Gif-sur-Yvette, France}

\author{Yannick Chassagneux}%
\affiliation{%
Laboratoire de Physique de l’Ecole normale supérieure, ENS, Université PSL, CNRS, Sorbonne Université, Université Paris Cité, F-75005 Paris, France\\
}%

\author{Cédric R. Mayer}
\affiliation{%
 Université Paris-Saclay, ENS Paris-Saclay, Centrale Supélec, CNRS, LuMIn (UMR 9024), Gif-sur-Yvette, France\\
}%

\author{Carole Diederichs}%
 \email{carole.diederichs@phys.ens.fr}
 
\affiliation{%
 Laboratoire de Physique de l’Ecole normale supérieure, ENS, Université PSL, CNRS, Sorbonne Université, Université Paris Cité, F-75005 Paris, France\\
}%
\alsoaffiliation{Institut Universitaire de France (IUF), 75231 Paris, France}

\title[An \textsf{achemso} demo]
  {Optical properties of single CsPbBr$_{3}$ perovskite quantum dots synthesized by a modified ligand-assisted reprecipitation method}

\begin{document}

\newpage

\begin{abstract}

Colloidal perovskite quantum dots (pQDs) are promising quantum light emitters, and investigations at the single pQD scale have so far relied mostly on hot-injection synthesis, which requires precise temperature control and an inert atmosphere. While alternative synthesis routes under milder conditions are often associated with structural and surface defects that may have limited impact in ensemble measurements, demonstrating high optical quality at the level of individual pQDs constitutes the most stringent benchmark for a new synthesis protocol. Here, we demonstrate that a modified ligand-assisted reprecipitation (LARP) approach yields CsPbBr$_3$ pQDs showing state-of-the-art optical properties at the scale of single emitters. By combining an amine-mediated post-synthetic size-trimming strategy with didodecyldimethylammonium bromide (DDAB) ligands for enhanced surface passivation and colloidal stability, we obtain isolated pQDs with stable emission and minimal spectral diffusion at cryogenic temperatures. Micro-photoluminescence experiments resolve the characteristic fine structure of the bright exciton, its low-energy optical phonon replicas, and the trion and biexciton states. Time-resolved and photon correlation measurements show a $\sim90$~\si{\pico\second} lifetime and high purity single photon emission, respectively. These results demonstrate that modified LARP synthesis constitutes an accessible alternative to hot injection, preserving the intrinsic excitonic and quantum optical properties of individual pQDs while offering greater flexibility for post-synthetic ligand engineering, as exemplified here by the use of DDAB for surface passivation.

\end{abstract}

\section{Introduction}

Colloidal lead halide perovskite quantum dots (pQDs) have rapidly emerged as a versatile class of nanomaterials for optoelectronics. They combine the appealing properties of bulk perovskites, characterized by direct band gaps, high absorption and defect tolerance, where shallow defects contribute marginally to non-radiative recombination, with the advantages of quantum confinement \cite{akkerman_review_2018}. Compared to their bulk or thin-film counterparts, pQDs thus offer fine tuning of their emission wavelength not only through compositional engineering but also via size control. Quantum confinement also results in enhanced photoluminescence quantum yield (PLQY) and spectrally narrow emission, which are key features for light emission applications \cite{sutherland_photonic_2016}. In particular, individual pQDs exhibit high purity single photon emission at room temperature, with limited spectral diffusion and blinking \cite{park_room_2015,raino_single_2016,zhu_room-temperature_2022}, highlighting their potential for quantum technologies. More recently, the observation of superfluorescence emission in self-assembled pQD superlattices~\cite{raino_superfluorescence_2018} has also stimulated active research on the materials engineering of pQD superstructures \cite{aneesha_halide_2025} and on the fundamental understanding of collective phenomena in pQDs \cite{levy_collective_2025}. These developments increasingly rely on precise knowledge of the intrinsic photophysical properties of pQDs at the individual scale.

To date, all single pQD studies have relied on the well-established hot-injection synthesis method \cite{protesescu_nanocrystals_2015}, which enables precise control over particle size and morphology but requires stringent temperature control (typically around 180~\si{\celsius}) and an inert atmosphere. In parallel, alternative synthesis routes under milder conditions have been developed, among which ligand-assisted reprecipitation (LARP) has emerged as an attractive room-temperature and ambiant atmosphere approach. Originally introduced for organic nanoparticles, such as polymeric and semiconducting nanostructures \cite{shamsi_metal_2019}, LARP was later adapted to hybrid organic-inorganic pQDs \cite{papavassiliou_nanocrystallinemicrocrystalline_2012} and optimized for MAPbX$_3$ (MA: methylammonium; X: Cl, Br or I) and CsPbBr$_3$ compounds, where solvent and ligand engineering significantly improved the PLQY \cite{zhang_brightly_2015, li_cspbx3_2016, wei_homogeneous_2017}, particularly for light-emitting devices \cite{du_high-quality_2017}. However, synthesis under softer conditions is often associated with degradation in the form of structural and surface defects. While such degradation may have limited impact in ensemble measurements, where optical properties are averaged over a large number of pQDs, it becomes critical at the level of individual emitters. Demonstrating high optical quality at the level of individual pQDs therefore sets a stringent benchmark for LARP synthesis protocols, which have so far only been investigated in ensemble studies. Recently, a modified LARP protocol employing amines as post-synthetic molecular scissors has been introduced to selectively trim pQDs and thus control their size and shape \cite{mayer_synthesis_2022}. This approach offers the possibility of combining the simplicity of room-temperature synthesis with improved structural control.

In this work, we synthesize CsPbBr$_3$ pQDs using this modified LARP protocol \cite{mayer_synthesis_2022}, further incorporating didodecyldimethylammonium bromide (DDAB) ligands to enhance surface passivation \cite{song_roomtemperature_2018} and to prevent aggregation during the dilution step required for single pQD studies. We investigate the intrinsic optical properties of these LARP-synthesized pQDs at the single emitter level by high-resolution micro-photoluminescence spectroscopy at cryogenic temperatures. We unambiguously identify key spectral signatures, including the neutral bright exciton and its characteristic fine structure splitting, as well as trion and biexciton states and optical phonon replicas. Furthermore, we demonstrate that individual LARP-synthesized pQDs exhibit single photon emission with pronounced photon antibunching under both continuous wave and pulsed excitation. Our results demonstrate that ligand-tailored LARP synthesis produces pQDs with intrinsic optical properties comparable to those of their hot-injection-synthesized counterparts, establishing it as an accessible synthesis route for both fundamental studies on individual pQDs and the development of perovskite-based quantum light sources.

\section{Results and discussion}

\begin{figure}[H]
  \centering
  \includegraphics{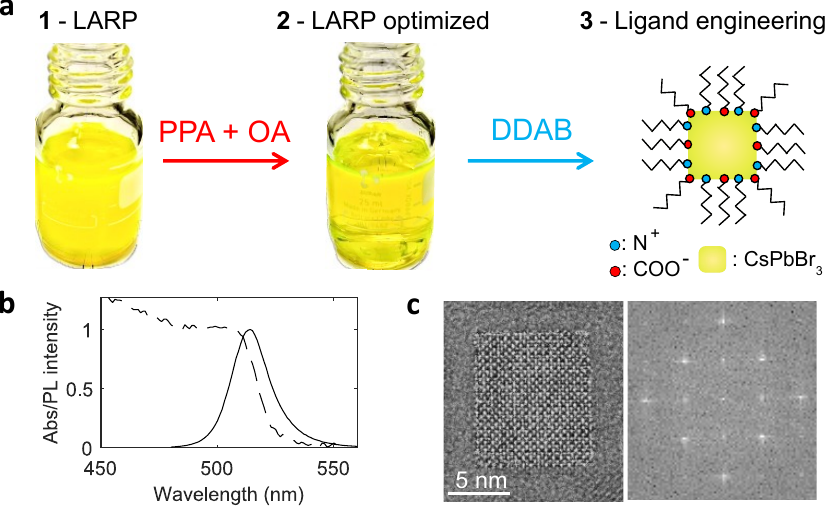}
  \caption{(a) Schematic illustration of the optimized LARP-based synthesis, outlining the critical steps used to obtain a monodisperse solution of surface-passivated CsPbBr$_3$ pQDs. Step~1: LARP synthesis, leading to a turbid solution of polydisperse pQDs. Step~2: Addition of PPA amines for post-synthetic size-trimming, leading to a transparent solution of monodisperse pQDs. Step~3: Addition of DDAB ligands, leading to a stable colloidal solution of surface-passivated pQDs. (b) Room-temperature absorption (dashed black line) of the resulting pQDs solution and its corresponding PL spectrum (solid black line) centered at 512~\si{\nano\meter} with a FWHM of 19~\si{\nano\meter}. (c) High resolution transmission electron microscopy (HRTEM) image of an individual CsPbBr$_3$ pQD (left) and the corresponding FFT-derived diffraction pattern (right).}
  \label{fig:fig1}
\end{figure}

The samples were prepared from a solution synthesized by LARP \cite{mayer_synthesis_2022}, where key optimizations were implemented to enable optical studies of individual CsPbBr$_3$ pQDs. Figure~\ref{fig:fig1}a summarizes the optimized procedure where all the steps were carried out under ambient conditions at room temperature. First, a conventional LARP synthesis using oleic acid (OA) as the sole ligand produces a turbid solution of polydisperse pQDs (Fig.~\ref{fig:fig1}a, step~1). To improve size and shape control, propylammonium (PPA) is then introduced, triggering a ligand-assisted size-trimming mechanism that yields a transparent dispersion (Fig.~\ref{fig:fig1}a, step~2) of monodisperse well-defined cubic pQDs (see SI for TEM characterization). At this stage, the pQDs surface is mostly coordinated at cationic sites (i.e. Cs$^+$, Pb$^{2+}$). Finally, didodecyldimethylammonium bromide (DDAB) is added to bind selectively to the bromide ions (Br$^-$), providing additional passivation and increasing the overall ligand coverage (Fig.~\ref{fig:fig1}a, step~3). This results in anionic and cationic passivation, effectively stabilizing the pQDs against aggregation and preserving their surface integrity throughout the subsequent dilution and deposition steps. The measured PLQY of the final solution is then equal to 94~\%. Figure~\ref{fig:fig1}b presents the absorption spectrum of the final solution together with the room-temperature ensemble PL spectrum, showing an emission centered at 512~\si{\nano\meter} with a Full Width at Half Maximum (FWHM) of 19~\si{\nano\meter} which is comparable to the room-temperature linewidths observed for individual hot-injection-synthesized pQDs \cite{park_room_2015,zhu_room-temperature_2022}. Such an ensemble PL linewidth indicates a very good size homogeneity of the LARP-synthesized pQDs, which is confirmed by the size distributions of $10.5 \pm 1.7$~\si{\nano\meter} retrieved from the TEM images (see SI). High-resolution TEM performed on the final solution further confirms the formation of cubic shaped pQDs. In particular, the HRTEM image in Figure~\ref{fig:fig1}c shows a representative single pQD with an edge length of 9.2~\si{\nano\meter} and a well-defined diffraction pattern in the Fourier space, demonstrating good crystalline quality and confirming the expected orthorhombic structure at room temperature \cite{mayer_2024_mLARP}. To perform optical studies of individual pQDs at cryogenic temperature, the pQDs solution is finally diluted in a 3~\% polystyrene (PS)-toluene mixture and spin-coated onto a silicon substrate. This results in a pQDs density of approximately $10^{-3}$~pQDs/\si{\micro\square\meter} embedded in a 150~\si{\nano\meter}-thick PS layer, which prevents degradation of the deposited pQDs \cite{raino_polymer_2019}. 

\begin{figure}[H]
  \centering
  \includegraphics{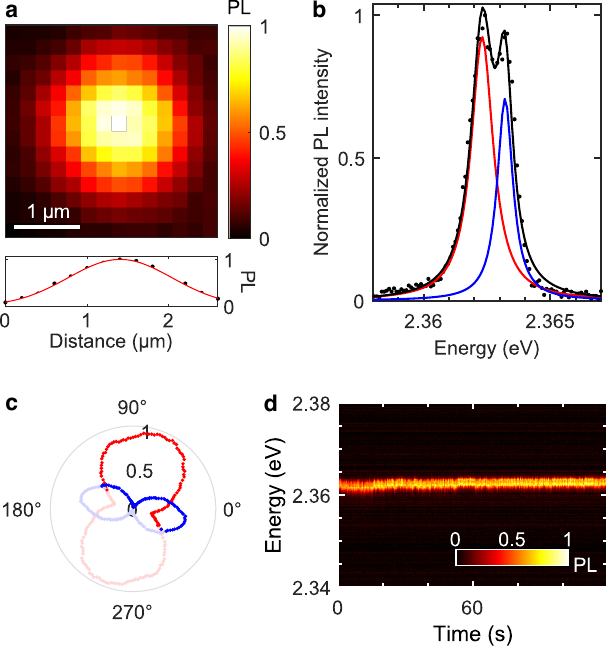}
  \caption{(a) High-resolution PL map of an isolated single pQD (distinct from the pQD under study), obtained with a laser spot size of $\sim$1.2~\si{\micro\meter} and a scanning step of 0.2~\si{\micro\meter}. Lower panel: horizontal line cut through the center of the map, fitted with a Gaussian profile (FWHM = 1.5~\si{\micro\meter}). (b) PL spectrum of the excitonic emission from a single pQD at 4~\si{\kelvin}. The black dots are experimental data, and the solid red and blue lines are Lorentzian fits with linewidths of 1.0~\si{\milli\electronvolt} and 0.7~\si{\milli\electronvolt}, respectively, separated by 0.9~\si{\milli\electronvolt}. (c) Emission polarization diagram, where the blue and red dots represent the intensity of each spectral component of the excitonic spectrum in (b). The faded points correspond to duplicated data used to extend the plot over the full 0-360\si{\degree} range. (d) Temporal PL trace recorded over 2~\si{\min} with a binning time of 200~\si{\milli\second}.}
  \label{fig:fig2}
\end{figure}

The results presented below summarize the optical response at 4~\si{\kelvin} of the same CsPbBr$_3$ pQD (unless otherwise specified) and are representative of measurements performed on other individual pQDs (see SI). First, to identify a well-isolated pQD prior to high-resolution spectral and temporal measurements, PL spatial maps are acquired using a scanning micro-PL setup, where a continuous wave (CW) excitation laser ($\lambda_{\mathrm{exc}} = 457$~\si{\nano\meter}) is focused on the sample through a microscope objective of numerical aperture N.A.=0.5 to a spot size of $\sim$1.2~\si{\micro\meter} and scanned across the surface using a fast steering mirror with a 0.2~\si{\micro\meter} step size. Figure \ref{fig:fig2}a presents a high-resolution PL map showing a well-defined symmetric emission spot, where a horizontal line cut through the center is fitted by a Gaussian with a FWHM of 1.5~\si{\micro\meter}. Such observation is a first signature of the emission from an isolated pQD. 

The PL spectrum associated to such PL spatial map is shown in Figure~\ref{fig:fig2}b. It consists of two Lorentzian lines separated by 0.9~\si{\milli\electronvolt}, with respective linewidths of 1.0~\si{\milli\electronvolt} and 0.7~\si{\milli\electronvolt}. The polarization diagram in Figure~\ref{fig:fig2}c further shows that these two lines are linearly and
orthogonally polarized. According to theoretical predictions for the band-edge exciton in inorganic pQDs \cite{becker_bright_2018,BenAich_bright_splitting_2019} and low-temperature experimental studies on CsPbBr$_3$ pQDs~\cite{raino_single_2016,fu_neutral_2017,ramade_FineStructureExcitons_2018,becker_bright_2018,amara_spectral_2023}, such energy splitting and polarization diagram are consistent with the expected bright exciton fine structure in the case of cubic shaped pQDs with aspect ratio close to one. In fact, at low temperature, the orthorhombic crystal symmetry should lift the degeneracy of the bright triplet excitonic state which can therefore be considered as three orthogonal emitting dipoles. Depending on their orientation relative to the detection axis and on the population of the emitting states, either two or three PL peaks are then typically resolved~\cite{amara_spectral_2023} (see SI for a three peak spectrum example). The spectra measured here are therefore fully consistent with the established fine structure signature of individual CsPbBr$_3$ pQDs. We note that the monoclinic cristal distortion reported at low temperature for LARP-synthesized pQDs \cite{mayer_synthesis_2022} may refine the detailed analysis but does not alter the overall interpretation of the exciton fine structure in the case of low crystal symmetry.

A crucial requirement for single quantum emitters is the stability of their emission over time. Figure \ref{fig:fig2}c presents a PL intensity time trace recorded for 2~\si{\min} with a 200~\si{\milli\second} binning time, showing no blinking behavior at this timescale and only limited spectral diffusion under moderate excitation power (below 10~\si{\micro\watt}), with no signs of degradation. For both components of the doublet, the standard deviation of the peak energy is approximately 300~\si{\micro\electronvolt}, and both lines are strongly correlated in energy fluctuations (see SI). This stability is likely due to the efficient surface passivation provided by the added DDAB ligands and the PS encapsulation.

\begin{figure} [H]
  \centering
  \includegraphics[width=\textwidth]{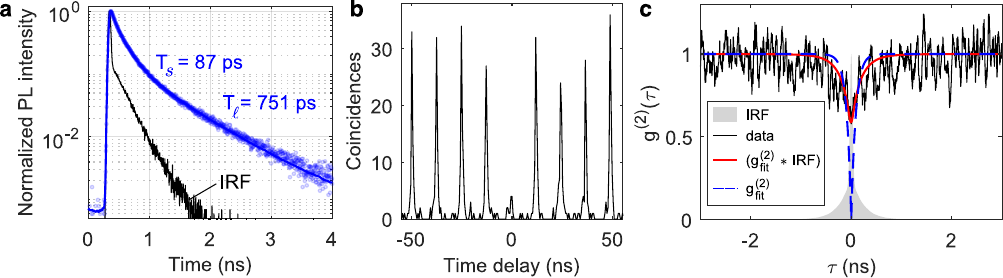}
  \caption{(a) Time-resolved PL of an individual pQD (blue dots). The black line is the Instrumental Response Function (IRF) measured at the pQD emission wavelength. The blue solid line is a bi-exponential fit of the PL decay, taking the IRF into account, with short- and long-time components $T_s\sim 87$~\si{\pico\second} and $T_\ell\sim 751$~\si{\pico\second}, respectively. (b) Intensity auto-correlation of the excitonic emission measured in a HBT setup under pulsed excitation, where $g^{(2)}(0)=0.12 \pm 0.06$ after background subtraction. (c) Same measurement as (b) under CW excitation, where the IRF of the HBT setup is shown as a grey shaded area and the experimental data (black line) are fitted by the theoretical $g_{\mathrm{fit}}^{(2)}$ function convoluted with the IRF (red line), giving the corrected value $g^{(2)}_{\mathrm{fit}}(0)=0.01 \pm 0.01$.}
  \label{fig:fig3}
\end{figure}

The emission dynamics and photon emission statistics are then investigated using time-resolved and photon correlation measurements based on Time-Correlated Single Photon Counting (TCSPC) techniques. Figure~\ref{fig:fig3}a shows the time-resolved PL decay, fitted by a convolution of the Instrumental Response
Function (IRF) measured at the pQD emission wavelength with a bi-exponential decay. A short-time component of $T_s\sim 87$~\si{\pico\second} and a long-time one of $T_\ell\sim 751$~\si{\pico\second}, the integrated weight of which is about 30~\%, are extracted. Such bi-exponential behavior is frequently reported for lead halide pQDs and has been attributed to the interplay between bright and dark exciton states \cite{tamarat_ground_dark_2019,amara_impact_2024}. The short-time component reflects the radiative recombination of the bright exciton (here all below 100~\si{\pico\second}), while the long-time contribution arises from thermal population exchange of the dark and bright states, followed by its subsequent recombination from the bright state. The observation of this slower channel therefore constitutes indirect evidence of the presence of a dark exciton state. Moreover, the ratio between the slow and the fast components suggests a sample temperature close to 20~\si{\kelvin}~\cite{amara_impact_2024}, rather than the expected 4~\si{\kelvin} in the cryostat. This deviation could originate from laser-induced heating or imperfect thermal interface of the sample holder.

\begin{figure}[H]
  \centering
  \includegraphics{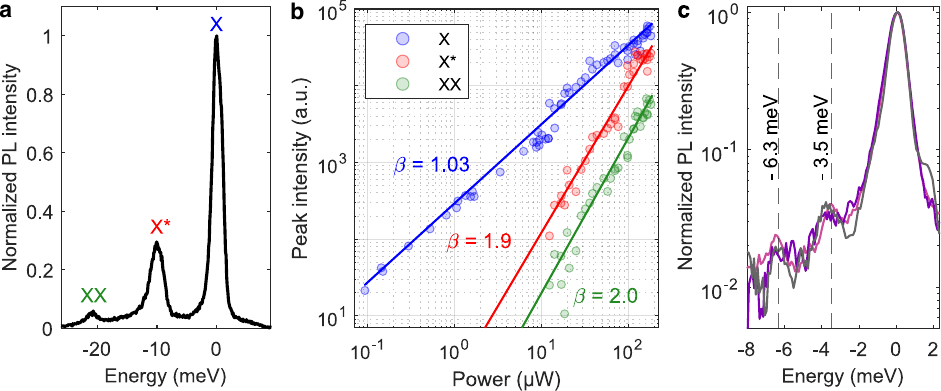}
  \caption{(a) PL spectrum of a single pQD showing the exciton (X), trion (X*), and biexciton (XX) emission lines. (b) Integrated PL intensity of each line as a function of excitation power plotted on a log-log scale. Experimental data (circles) are fitted with power-law dependencies (solid lines), from which the exponent $\beta$ is extracted and indicated next to each fit. (c) PL spectra of three different pQDs, including the pQD analyzed throughout this work, plotted on a semi-logarithmic scale. Two optical phonon replicas of the excitonic emission are observed at $-3.5$~\si{\milli\electronvolt} and $-6.3$~\si{\milli\electronvolt}, highlighted by vertical dashed lines.}
  \label{fig:fig4}
\end{figure}

To assess the photon emission statistics of the pQD, the second-order intensity correlation function $g^{(2)}(\tau)$ is measured using a Hanbury Brown and Twiss (HBT) setup. The photon arrival times recorded at each output port of a beamsplitter in the detection path are used to construct the coincidences histogram under pulsed excitation (Fig.~\ref{fig:fig3}b) and the normalized intensity autocorrelation function under CW excitation (Fig.~\ref{fig:fig3}c). In both configurations, the excitation power is kept low enough to maintain an average exciton population below 0.1, ensuring operation in the linear excitation regime. Moreover, the exciton line is spectrally filtered to suppress any biexciton or trion contributions (see Fig.~\ref{fig:fig4}). Under pulsed excitation (Fig.~\ref{fig:fig3}b), the coincidence peak at zero time delay is strongly suppressed, with $g^{(2)}(0)=0.12 \pm 0.06$ after background subtraction. This pronounced photon antibunching, with $g^{(2)}(0) < 0.5$, is a clear signature of single photon emission and further confirms that single pQDs synthesized by the modified LARP protocol can be used as quantum emitters. Under CW excitation (Fig.~\ref{fig:fig3}c), a dip at zero delay is also visible, but its resolution is strongly limited by the finite detector response, which is comparable to the exciton emission lifetime and therefore leads to an overestimated raw value of $g^{(2)}(0)$. To retrieve the intrinsic photon statistics, the data is fitted to the convolution of the IRF with the theoretical second-order correlation function at low pumping rate, $ g_{\mathrm{fit}}^{(2)}(\tau)= 1-A e^{- |\tau| /T_s}$, where $T_s$ is fixed to the exciton lifetime measured from time-resolved PL and $A$ quantifies the single photon purity. This fitting procedure allows us to deduce the intrinsic value of $g_{\mathrm{fit}}^{(2)}(0)=0.01 \pm 0.01$, confirming high-purity single photon emission.

Beyond the excitonic emission, the PL spectrum of a single pQD at cryogenic temperature can exhibit additional features. As shown in Figure~\ref{fig:fig4}a, two extra red shifted lines appear at energy shifts of approximately 10 and 21~\si{\milli\electronvolt} from the exciton when the excitation power is increased. These energy shifts are consistent with previously reported binding energies of biexciton and charged exciton states in CsPbBr$_3$ pQDs of comparable size \cite{amara_spectral_2023}. To further characterize these features, the integrated intensity of each spectral component is plotted as a function of excitation power (Fig.~\ref{fig:fig4}b). At low excitation power (below 10~\si{\micro\watt}), only the exciton line is observed. As the power increases, the probability of creating excitonic complexes rises and the red shifted lines progressively appear. Their power dependence follows characteristic scaling laws that can be described within the random population model~\cite{grundmann_theory_1997}, which assumes a Poisson distribution for the population of any excitonic complex. In this framework, the probability of having $n$ excitons, and thus the intensity of the corresponding emission line, is given by $n = \langle n\rangle e^{-\langle n\rangle}$, where the mean number of excitonic complex $\langle n\rangle=P/P_{sat}$, with $P$ the excitation power and $P_{sat}$ the saturation power. All measurements are performed below saturation ($P\lesssim P_\mathrm{sat}$), as confirmed by the absence of slope changes in the log-log plot in Figure \ref{fig:fig4}c. In this low-power regime, the model simplifies to $ \langle n\rangle = \alpha P_{exc} ^{\beta}$, where $\alpha$ is a scaling factor and the exponent $\beta $ reflects the excitonic nature of the emitting state. Fitting the experimental data yields $\beta = 1.03 \pm 0.03$, $\beta = 1.9 \pm 0.1$ and $\beta = 2.0 \pm 0.2$ from the highest- to the lowest-energy spectral contributions, respectively. These values allow the assignment of the corresponding lines to the neutral exciton (X), the charged exciton or trion (X*), and the biexciton (XX). The superlinear scaling of the trion emission further indicates that the additional charge is photo-generated under optical excitation rather than permanently trapped in the pQD \cite{abbarchi_poissonian_2009}. We note that the exponents associated with the trion and biexciton emissions overlap within their uncertainties, which could hinder their identification. However, the above-mentioned energy shifts with respect to the neutral exciton corroborate their assignment.

Finally, Figure~\ref{fig:fig4}c presents highly resolved PL spectra of the neutral exciton from three different pQDs plotted on a semi-logarithmic scale. In all cases, weak additional features are observed on the low-energy side of the exciton at similar energy shifts of approximately 3.5 and 6.3~\si{\milli\electronvolt}. These shifts match the first transverse optical phonon modes reported in CsPbBr$_{3}$, allowing these features to be assigned to exciton phonon replicas~\cite{fu_neutral_2017,cho_excitonphonon_2022,amara_spectral_2023}. The relative intensity of these replicas with respect to the exciton zero phonon line (ZPL) enables the extraction of the Huang-Rhys factor, which quantifies the strength of exciton-phonon coupling. We obtain a value of 0.01, showing that the emission originates predominantly from the ZPL (see SI). We note that in the previous power study, to ensure accurate extraction of the integrated intensities shown in Figure \ref{fig:fig4}b, each spectrum was fitted using a set of Lorentzian functions accounting for both the ZPL and the associated phonon replicas. We verified that including or excluding the phonon replicas in the fitting procedure does not affect the extracted power-law exponents, demonstrating that the phonon sidebands scale proportionally with the corresponding ZPL (see SI). This behavior reflects their common origin in excitonic recombination process, with a relative weight governed by the exciton-phonon coupling rather than by the excitation density.

\subsection{Outline}

In summary, we have demonstrated that single CsPbBr$_3$ pQDs synthesized using a modified ligand-assisted reprecipitation method, combined with DDAB ligand engineering, exhibit intrinsic optical properties at cryogenic temperature comparable to those of pQDs synthesized by hot injection. At the level of individual pQDs, we observe the characteristic spectral signatures of the bright exciton, including its linearly-polarized fine structure and optical phonon replicas, as well as the multiexcitonic trion and biexciton states. Furthermore, we demonstrated that LARP-synthesized pQDs exhibit fast radiative and high-purity single photon emission, similar to more conventional pQDs. By meeting the stringent requirements imposed by single emitter spectroscopy, these results establish the modified LARP approach as a well-controlled alternative to hot-injection synthesis for the production of high quality single emitters. The versatility of this room-temperature protocol further provides opportunities for tailored post-synthetic ligand engineering which could be crucial for assembling pQDs into optimized superstructures and exploring collective quantum emission phenomena.

\section{Experimental}

\subsection{Chemicals} 

Cesium carbonate (Cs$_2$CO$_3$, $\ge 99.9$~\%), lead(II) bromide (PbBr$_2$, $\ge 98$~\%), oleic acid (OA, technical grade, 90~\%), tetraoctylammonium bromide (TOABr, $\ge 99$~\%), N,N-dimethylformamide (DMF, anhydrous, 99.8~\%), tetrahydrofuran (THF, $>97$~\%), 3-phenylpropylamine (PPA, C$_9$H$_13$N, $>97$~\%), didodecyldimethylammonium bromide (DDAB, $\ge 98$~\%), methyl acetate ($\ge 97$~\%) and toluene (anhydrous, 99.8~\%) were purchased from commercial suppliers and used without further purification. All solvents were handled under ambient conditions in a fume hood unless otherwise stated.

\subsection{Synthesis}

CsPbBr$_3$ nanocrystals were synthesized using a modified ligand-assisted reprecipitation (LARP) method. Two precursor solutions were prepared separately.
Lead precursor solution: PbBr$_2$ (0.1~\si{\milli\mol}) and TOABr (0.2~\si{\milli\mol}) were dissolved in 10~\si{\milli\liter} of anhydrous toluene under heating ($<90$~\si{\celsius}) with continuous stirring to accelerate dissolution; the solution was subsequently cooled to 1–2~\si{\celsius}. Cesium precursor solution:
Cs$_2$CO$_3$ (0.1~\si{\milli\mol}) was dissolved in 2~\si{\milli\liter} of OA under heating until complete dissolution and then allowed to cool to room temperature.
The two precursor solutions were rapidly combined under vigorous stirring at room temperature. Subsequently, 1.125~\si{\milli\liter} of THF was added to facilitate nanocrystal formation and assist in subsequent purification.

\subsection{Purification of the CsPbBr$_3$ nanocrystals}

The resulting nanocrystal dispersion was centrifuged at 10,000~rpm for 20~\si{\min} to remove unreacted precursors and excess ligands. The supernatant was discarded, and the precipitated nanocrystals were redispersed in toluene for further processing.

\subsection{Size tuning and stability optimization}

A trimming solution was prepared by dissolving Cs$_2$CO$_3$ (8~\si{\milli\gram}), PbBr$_2$ (36~\si{\milli\gram}), and PPA (175~\si{\micro\liter}) in DMF (2.2~\si{\milli\liter}). Subsequently, 320~\si{\micro\liter} of the cutting solution was added to the purified nanocrystal dispersion under stirring. After 5~\si{\min}, 1.73~\si{\milli\liter} of DDAB (10~\si{\milli\gram}/\si{\milli\liter} in solvent) was added to enhance nanocrystal stability. If required, the resulting solution was further purified by centrifugation using a 1:1 (v/v) mixture of methyl acetate and the perovskite nanocrystal solution.

\subsection{Sample preparation}

The nanocrystal solution obtained via the modified LARP method was diluted by a factor of 40,000 in two successive steps using toluene containing 3~\% (w/w) polystyrene to protect the nanocrystals from ambient air, and 2~\% (w/w) oleic acid to ensure the stability of the surface ligand shell. Thin films were subsequently deposited by spin coating. The substrates were covered with the diluted solution and spun at 3,000~rpm for 60~\si{\sec}, resulting in the formation of homogeneous quantum dot films with a surface density of approximately $10^{-3}$~pQDs/\si{\micro\square\meter}.

\subsection{Characterization}

The optical properties of the CsPbBr$_3$ nanocrystals in solution were characterised by a spectrofluorometer, including the UV-Vis absorption and photoluminescence spectra and the PLQY at room temperature. Morphology and size distribution were examined by transmission electron microscopy (TEM).

\subsection{Optical studies}

All measurements were performed in a home-built micro-PL setup with the sample mounted inside a liquid helium flow cryostat. Excitation and collection were performed through a microscope objective of numerical aperture N.A.=0.5. For steady state PL measurements, the sample was excited by a laser diode at $\lambda = 457$~\si{\nano\meter}. The PL was then separated from the excitation laser using a dichroic mirror and a long-pass filter, then dispersed by a grating monochromator and detected with a Peltier-cooled CCD camera for spectral studies. For photon correlation and PL lifetime measurements, the sample was excited by a frequency-doubled Ti:Sa laser ($\lambda=435$~\si{\nano\meter}, 100~\si{\femto\second} pulse duration, 80~\si{\mega\hertz} repetition rate). The PL was then filtered by a tunable 1~\si{\nano\meter} bandpass filter and detected using a time-correlated single photon counting (TCSPC) system with avalanche photodiodes of 35~\si{\pico\second} time response.

\begin{acknowledgement}

This work was supported by the French National Research Agency (ANR) through the project CSUPER2 (ANR-24-CE09-3415). The authors thank Elsa Cassette, Emmanuelle Deleporte, Jean-Sébastien Lauret, and Gaëlle Trippé-Allard for the fruitful discussions.

\end{acknowledgement}

\begin{suppinfo}

Provided after the references: supplemental figures and further details on TEM characterization, spectral features from other individual pQDs, PL temporal stability, polarization analysis, power dependence, and Huang-Rhys factors.

\end{suppinfo}

\bibliography{pQD_LARP_biblio}


\setcounter{section}{0}
\setcounter{figure}{0} 

\renewcommand{\thefigure}{S\arabic{figure}}

\vspace*{5cm}

\section{\centering Supplementary information for:} 

\vspace*{2cm}

\section{\centering Optical properties of single CsPbBr$_{3}$ perovskite quantum dots synthesized by a modified ligand-assisted reprecipitation method} 

\newpage

\subsection{Transmission electron microscopy characterization}

\begin{figure}[H]
    \centering
    \includegraphics{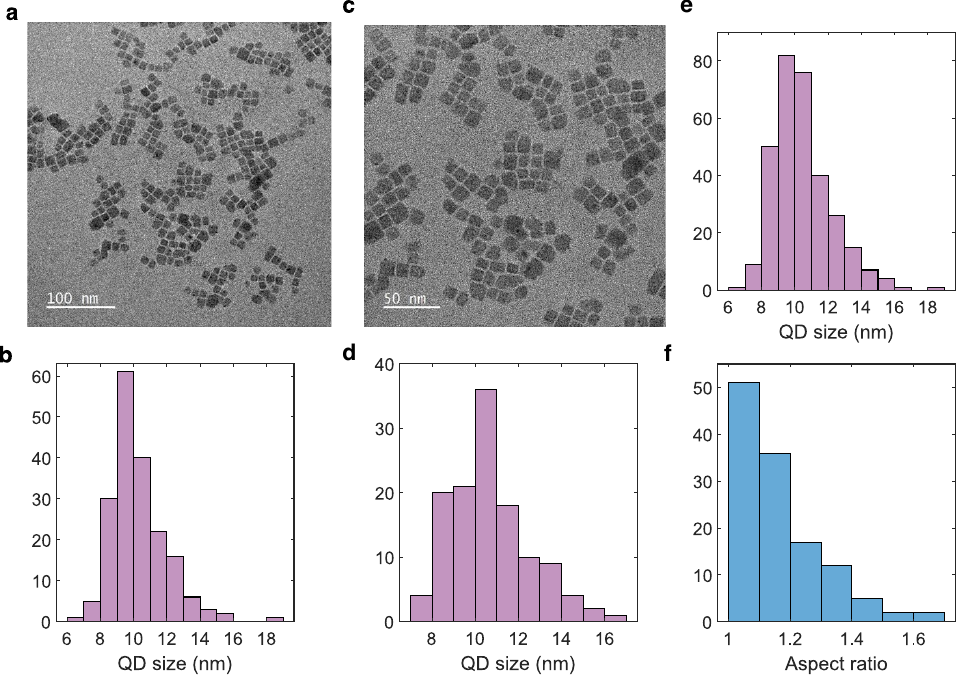}
    \caption{(a,c) TEM images of the CsPbBr$_3$ pQDs synthesized by the optimized LARP synthesis protocol.  (b,d) Corresponding particle size distribution histograms with average length of $10.3  \pm 1.7$~\si{\nano\meter} and $10.7 \pm 1.8$~\si{\nano\meter}, respectively. (e) Total size distribution obtained by summing histograms (b) and (d). (f) Aspect ratio extracted from the TEM images, with a mean value of  $1.17  \pm 0.15$.}
    \label{fig:histo_TEM}
\end{figure}

The dimensions of the pQDs were extracted from the TEM images in Figure~\ref{fig:histo_TEM}a and Figure~\ref{fig:histo_TEM}c by taking the mean of the width and height of each individual pQD. Figure~\ref{fig:histo_TEM}b and Figure~\ref{fig:histo_TEM}d show histograms of the measured pQD sizes, with average sizes of $10.3  \pm 1.7$~\si{\nano\meter} and $10.7 \pm 1.8$~\si{\nano\meter}, respectively. The mean aspect ratio is 1.17, with a standard deviation of 0.15, based on the overall size measurements.

\subsection{Size-energy comparison}

\begin{figure}[H]
    \centering
    \includegraphics{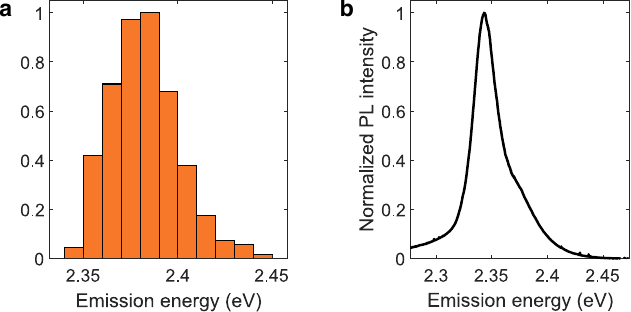}
    \caption{(a) Emission energy histogram deduced from the pQDs sizes measured in Fig.~\ref{fig:histo_TEM}e using the size-energy correspondence of Ref.~\citenum{amara_spectral_2023}, showing an average energy of $2.38 \pm 0.02$~\si{\electronvolt}. (b) PL spectrum at 4~\si{\kelvin} of an ensemble of pQDs deposited in polystyrene, centered at 2.34~\si{\electronvolt} with a FWHM of $\sim 0.03$~\si{\electronvolt}.}
    \label{fig:histo_energy_PL}
\end{figure}

Figure~\ref{fig:histo_energy_PL}a presents the emission energy distribution of the pQDs with a standard deviation of 18~\si{\milli\electronvolt}, obtained from the size distribution shown in Figure~\ref{fig:histo_TEM}e using the size-energy correspondence reported by Amara \textit{et al.} \cite{amara_spectral_2023}. Figure~\ref{fig:histo_energy_PL}b presents a PL spectrum of a pQDs ensemble at cryogenic temperature, with a FWHM of about 30~\si{\milli\electronvolt}. A slight shift in the central energy is observed between the PL spectrum and the energy histogram, likely due to the inherent uncertainties in the empirical extraction of the formula parameters and the differences in measurement temperatures and reference conditions. Nevertheless, the comparison suggests that the size distributions derived from TEM are consistent with the main features of the ensemble PL spectrum.

\subsection{Spectral features from various pQDs}

\begin{figure}[H]
    \centering
    \includegraphics{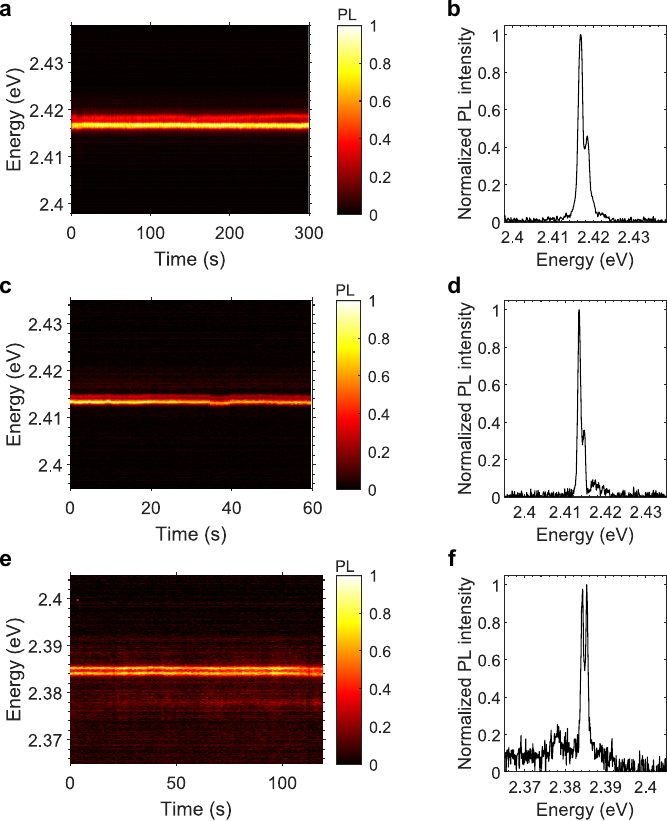}
    \caption{Temporal photoluminescence traces (a, c, e) and corresponding time-integrated PL spectra (b, d, f) for three different single pQDs.}
    \label{fig:3pQDs}
\end{figure}

To illustrate the reproducibility of the measurements discussed in the main text, we present temporal PL traces (Figure~\ref{fig:3pQDs} a, c, e) and corresponding spectra (Figure~\ref{fig:3pQDs} b, d, f) from three different single pQDs under CW laser excitation. For the first two pQDs, measurements were performed at a power of 450~\si{\nano\watt} with an integration time of 0.5~\si{\second} per time bin. For the third pQD, a slightly higher power of 780~\si{\nano\watt} and an integration time of 1~\si{\second} were used.

\subsection{Temporal stability of the PL emission}

\begin{figure}[H]
    \centering
    \includegraphics{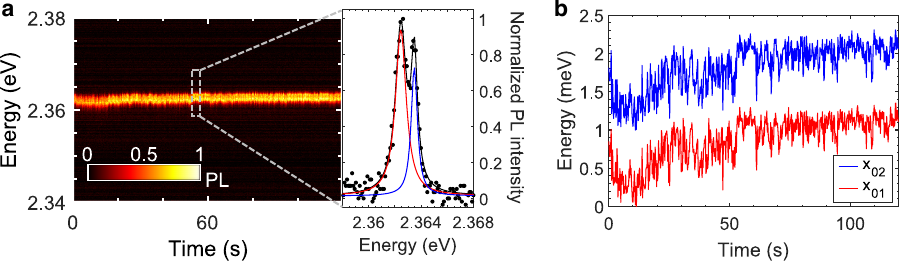}
    \caption{(a) (left) Temporal PL trace of a single pQD recorded over 2~\si{\min} with a binning time of 200~\si{\milli\second}. Close-up on one time bin of the temporal trace, showing the corresponding PL spectrum (right) characterized by a doublet. Black dots are experimental data, and the solid red and blue lines are Lorentzian fits with linewidths of 1.0~\si{\milli\electronvolt} and 0.7~\si{\milli\electronvolt}, respectively, separated by 0.4~\si{\milli\electronvolt}. (b) Central energy of each lorentzian peak ($x_{01}$ and $x_{02}$) extracted from the fit of every spectrum in the temporal trace, plotted as a function of time.}
    \label{fig:figSI_zoom_trace}
\end{figure}

To further characterize the temporal stability of the excitonic emission, we analyze the time-resolved PL trace shown in Figure~\ref{fig:figSI_zoom_trace}a (left). The PL signal was recorded continuously over 2~\si{\minute} with a binning time of 200~\si{\milli\second}. For each time bin, the corresponding PL spectrum was extracted and fitted with a sum of two Lorentzian functions, accounting for the two components of the excitonic doublet. A representative spectrum from a single time bin is shown in Figure~\ref{fig:figSI_zoom_trace}a (right), where both peaks are well described by Lorentzian lineshapes. The extracted linewidths remain narrow (below 1.0~\si{\milli\electronvolt}) and stable over time.
The central energies of both peaks were then tracked over the entire temporal trace, as shown in Figure~\ref{fig:figSI_zoom_trace}b. While small energy fluctuations ($\sim1$~\si{\milli\electronvolt}) are observed, both components of the doublet evolve synchronously: we find a strong correlation coefficient between the two peak energies of 0.99. Such a high degree of correlation further indicates that the two lines originate from the same emitter and experience the same local electrostatic environment. 

\subsection{Exciton fine structure: polarization analysis of a triplet}

\begin{figure}[H]
    \centering
    \includegraphics{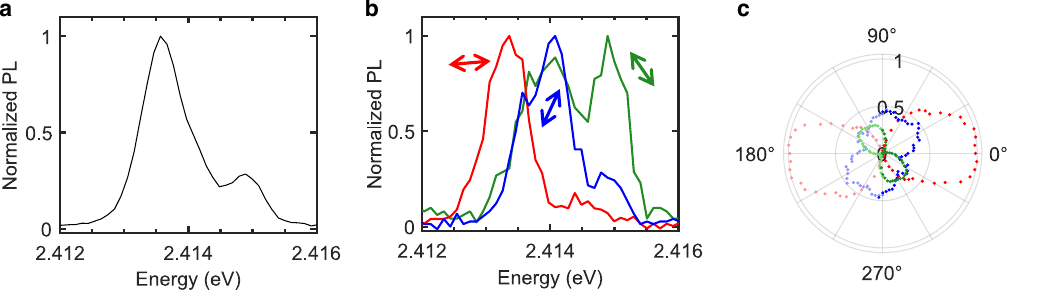}
    \caption{(a) PL spectrum of a single pQD exhibiting an exciton triplet emission. (b) Polarization-resolved PL spectra recorded at three polarization angles maximizing each individual emission line. (c) Corresponding polarization diagram where each of the three lines is linearly polarized. The faded points correspond to duplicated data used to extend the plot over the full 0-360\si{\degree} range.}
    \label{fig:trois-raies}
\end{figure}

In addition to the doublet spectra presented in Fig.~\ref{fig:3pQDs} and in the main text, a triplet can be observed in the PL spectrum of some pQDs, as shown in Figure \ref{fig:trois-raies}. Depending on the line splittings and relative intensities, the individual line contributions can be hidden in the unpolarized spectrum (see Fig.~\ref{fig:trois-raies}a). Polarization measurements are then performed to better resolve each component. Figure~\ref{fig:trois-raies}b presents three spectra recorded at polarization angles that maximize each of the three emission lines. Full polarization measurements were analyzed by fitting each spectrum with a sum of Lorentzian functions. The integrated intensity of each peak was extracted as a function of the polarization angle to obtain the polarization diagram shown in Figure~\ref{fig:trois-raies}c. As expected for the excitonic fine structure, each emission line exhibits linearly polarized emission.

\subsection{PL spectrum power dependence}

\begin{figure}[H]
    \centering
    \includegraphics{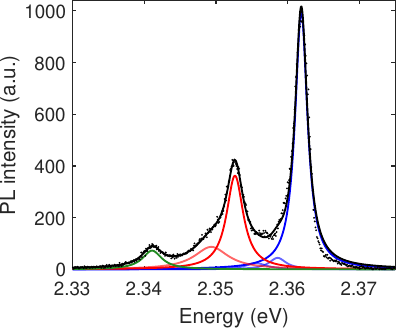}
    \caption{PL spectrum of a single CsPbBr$_3$ pQD at high excitation power (128 µW), together with the Lorentzian fits used for extracting the integrated intensities. The main zero-phonon lines (ZPL) of the exciton (X, blue), trion (X*, red), and biexciton (XX, green) are fitted individually. In addition, two phonon replicas of X at energy shifts of 3.3~\si{\milli\electronvolt} and 6.3~\si{\milli\electronvolt} from the ZPL, as well as one phonon replica of X* at 3.3~\si{\milli\electronvolt}, are fitted by shaded solid lines. The black dots are experimental data and the solid black line is a fit of the total spectrum.}
    \label{fig:PL_replica}
\end{figure}

To accurately determine the excitation power dependence of the different emission lines of a pQD, the PL spectra at high excitation power were systematically fitted using a sum of Lorentzian functions as illustrated in Figure~\ref{fig:PL_replica}. The fits include the zero-phonon lines (ZPLs) of the exciton (X), trion (X*), and biexciton (XX), together with their phonon replicas when present. Here, the X line exhibits phonon sidebands at energy shifts of 3.3~\si{\milli\electronvolt} and 6.3~\si{\milli\electronvolt}, while X* shows a phonon replica at 3.3~\si{\milli\electronvolt} as expected from previous studies \cite{amara_spectral_2023}. The inclusion of these replicas ensures that the integrated intensity of each excitonic transition is not underestimated at high excitation powers, where phonon-assisted emission becomes significant. 

\subsection{Evaluation of the Huang-Rhys factors}

\begin{figure}[H]
    \centering
    \includegraphics{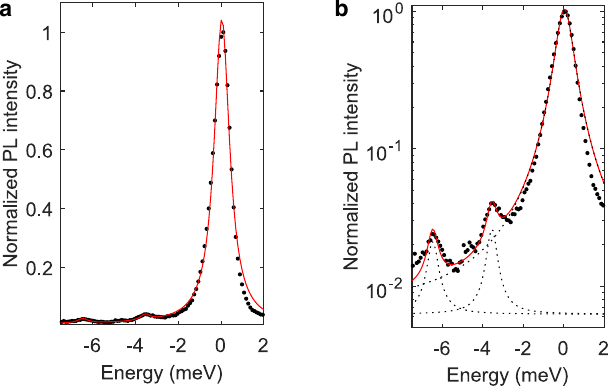}
    \caption{PL spectrum of a single pQD plotted on a linear (a) and semi-logarithmic scale (b). Two optical phonon replicas of the excitonic emission are observed on the low-energy side of the ZPL. Black dots correspond to experimental data, and the red solid line to the global fit, composed of three Lorentzian contributions (black dashed lines).}
    \label{fig:fig_HR}
\end{figure}

Figure~\ref{fig:fig_HR}a presents a single pQD PL spectrum where two optical phonon replicas $R_j$ ($j=1,2$) associated with the exciton emission can be resolved. These replicas are weak but resolved in the semi-logarithmic representation in Figure~\ref{fig:fig_HR}b. The spectrum is fitted using a sum of Lorentzian functions accounting for the exciton ZPL and its phonon replicas such as in Ref.~\citenum{cho_excitonphonon_2022}:

\begin{equation}
    I(E)=I_X(E)+\sum_{j=1} ^2 I_{R_j}(E), \quad I_{i}(E)= \frac{A_i}{1+\left(\dfrac{E-E_i}{\Gamma_i/2}\right)^2}
\end{equation}

\noindent where $A_i$, $E_i$, and $\Gamma_i$ are the amplitude, central energy, and width of each contribution, respectively. To assess the strength of the exciton-phonon coupling, we extract the Huang-Rhys factor $S_j$ for each optical phonon mode $j$, defined as the ratio of the integrated intensity of the $j$th phonon replica to that of the ZPL, leading to:
\begin{equation}
   S_{R_j} =\frac{ A_{R_j} \Gamma_{R_j}}{ A_{X} \Gamma_{X} }, \quad j = 1,2 
\end{equation}

\noindent We obtain $S_1 \approx 0.01$ and $S_2 \approx 0.01$, indicating very weak exciton-phonon coupling for both optical phonon modes.

\end{document}